\documentstyle[preprint,epsfig,11pt]{article}

\newcommand{\be}[1]{\begin{equation} \label{(#1)}}
\newcommand{\ee}{\end{equation}}
\newcommand{\ba}[1]{\begin{eqnarray} \label{(#1)}}
\newcommand{\ea}{\end{eqnarray}}

\begin{document}

\title{\Large{\bf The Parton-Hadron Phase Transition in Central Nuclear Collisions
 at the CERN SPS $^*$}}
\maketitle

\begin{center}
Reinhard Stock, Department of Physics, Frankfurt University
\end{center}

\abstract
\noindent
{A selection of recent data referring to $Pb+Pb$ collisions at the SPS CERN
energy of $158\:GeV$ per nucleon is presented which might describe the
state of highly excited strongly interacting matter both above and
below the deconfinement to hadronization (phase) transition predicted by
lattice  QCD. A tentative picture emerges in which a partonic state
is indeed formed in central $Pb+Pb$ collisions which hadronizes at about
$T=185\:MeV$, and expands its volume more than tenfold, cooling to about
$120\:MeV$ before hadronic collisions cease. We suggest further that all
SPS collisions, from central $S+S$ onward, reach that partonic phase,
the maximum energy density increasing with more massive collision systems.}

\section{Relativistic Nuclear Collisions}
Astrophysical objects and processes, both connected with very early and very
late phenomena in the cosmological evolution of strongly interacting matter,
present an enormous challenge to modern nuclear and particle physics: we can
recreate the conditions prevailing during the late nanosecond era of the
cosmological expansion (when free quarks and gluons hadronized to isolated
protons and neutrons), or during the late stages of a violent supernova
stellar implosion (when the properties of highly compressed nuclear matter
decide the final avenue leading either into a superdense neutron star or into
a black hole) in experiments carried out in the terrestrial laboratory, by
colli\-ding heavy nuclei at relativistic energy.
These studies culminate, for the time being, in the CERN SPS $^{208}Pb$
beam facility which accelerates $Pb$ nuclei to $158\: GeV$ per nucleon
(about $33\: TeV$ total energy). Ongoing programs at BNL and CERN will
vastly extend the energy domain from $ \sqrt{s} \approx 17\: GeV$ at the SPS to
collider mode experiments with $ \sqrt{s} = 200\: GeV$ (RHIC) and
$\sqrt{s} \approx 5\: TeV$ (LHC).

\noindent
The common idea of these investigations is to create extended "fireball"
volumes of strongly interacting matter in head-on collisions of heavy nuclei,
creating an average energy density reaching (at the SPS) or far exceeding
(at RHIC and LHC) the "critical" value of about $1.5\: GeV/fm^3$, at which
modern Lattice QCD theory predicts a sudden departure, concerning the
specific heat, the number density of degrees of freedom, the constituent
quark mass scale etc., away from the expected behaviour of a densely packed
liquid of hadrons. Does the hadron degree of freedom melt away at this
transition point, giving rise to a continuous
QCD state in which massive "constituent" quarks turn into nearly massless
"current" (QCD) quarks - a phenomenon associated with the concept of
chiral symmetry restoration - and in which colour carrying partons acquire
a finite mobility, i.e. they approach deconfinement in an extended plasma
state of {\it nuclear} dimension, i.e. $10\:fm$, large in comparison
to typical confinement scales of about $1\:fm$.

\begin{figure}
\hspace{2cm}
\begin{minipage}[t]{5cm}
\epsfig{figure=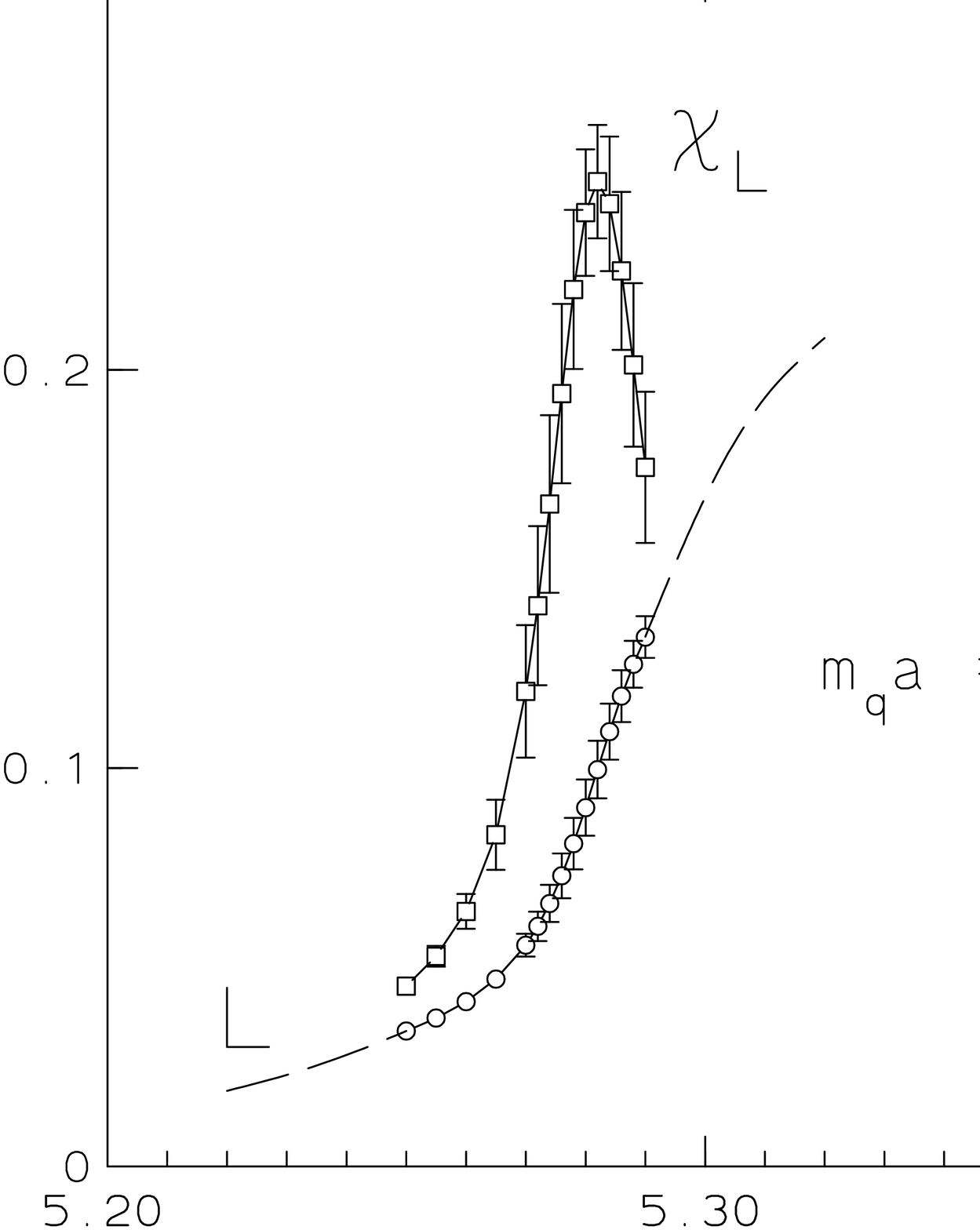,height=5cm, width=4cm}
\end{minipage}
\hspace{0,5cm}
\begin{minipage}[t]{5cm}
\epsfig{figure=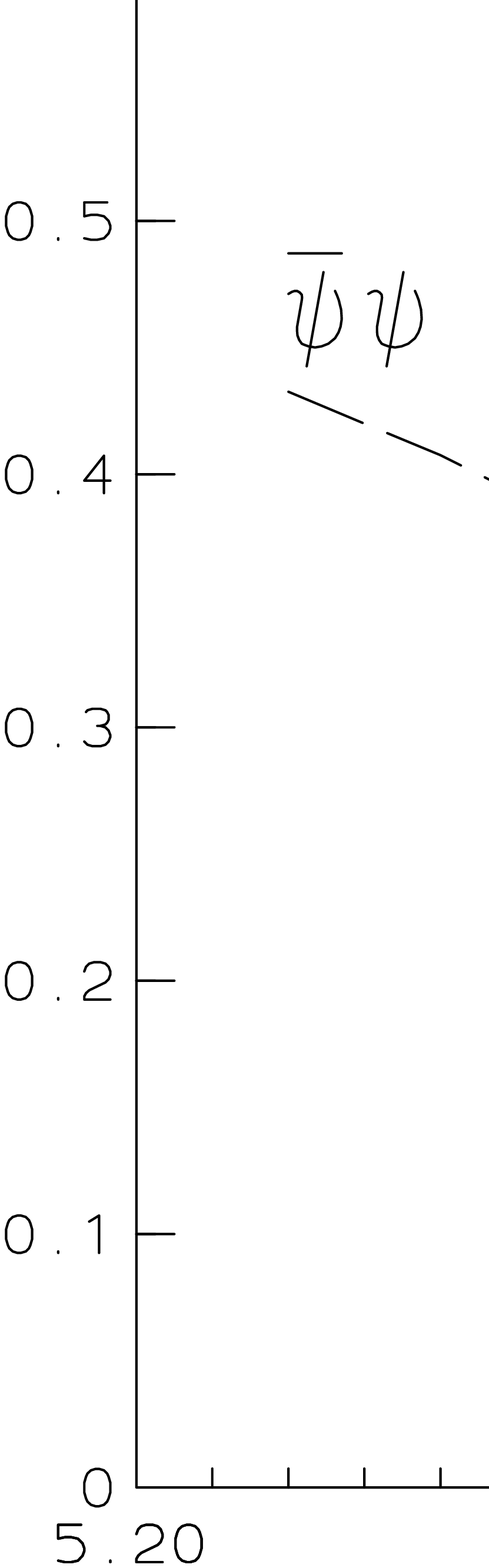,height=5cm, width=4cm}
\end{minipage}
\caption{Lattice QCD results for the Wilson loop L and the quark
mass scale $\langle \overline{\Psi}\Psi \rangle$ vs. $6/g^2$ with $g$
the lattice coupling constant.}
\end{figure}

\noindent
The present state of the art in lattice QCD finite temperature
calculations \cite{1} is illustrated in $Fig.~1$. On an $8^3\:\times\:4$
lattice with two dynamical quark flavours, the Wilson loop $L$ and the
effective quark mass scale $\langle\overline{\Psi}\Psi\rangle$ are
calculated along with the corresponding
generalized susceptibilities. The quantity $L$ depends on the free quark
energy, $L \approx exp [-F/T]$, and can be understood as a measure of
colour mobility vs. colour confinement. A sharp jump occurs as the temperature
given in units of the inverse lattice coupling constant $g$ reaches a
critical value (note the very narrow scale), suggesting a phase change or a
phase transition. At exactly the same position the quark mass (the "quark
condensate") drops steeply. This suggests that quarks and, hence,
hadrons loose their mass at a critical temperature $T_c$,
 and simultaneously acquire a finite free ener\-gy
in the medium, resulting in a finite mobility which indicates deconfinement.
This interpretation is supported by a concurrent steep jump in the energy
density $E/T^4$
(not illustrated). We thus expect a new QCD phase of matter setting in
at $T_c=160-200\:MeV$ which exhibits a critical energy density
of 1.5 to 3.0 $GeV/fm^3$.

\noindent
How to search experimentally for this new phase? In the following
chapters I will consider some of the appropriate physics
quantities,
 and present relevant data stemming from central
collisions of SPS sulphur ($^{32}S$) and lead ($^{208}Pb$) beam
projectiles with various nuclear targets. At first
we may ask whether  the critical energy density is reached (or even
surpassed) in the primordial interaction volume. The conclusion is
affirmative  as will be described in Section 2. The interaction
volume may reach energy densities characteristic of a partonic,
extended system. The immediate next question concerns observables
that respond directly to a transient, deconfined state (ideally a
quark-gluon-plasma state): Section 3 establishes the $J/\Psi$
production yield as one of the most informative, relevant
observables, concluding that the data appear to be
compatible with the hypothesis of deconfinement reached in central
$Pb+Pb$ collisions. The suppression of the $J/\Psi$ yield,
however, answers merely the "to be or not to be" question of
deconfinement. Supposing that a deconfined, partonic state was
indeed created in the primordial reaction volume, at the maximal
energy density attainable in CERN SPS collisions, the next equally
important question concerns observables elucidating the nature of
the parton to hadron phase transition occurring
 in the dynamical evolution during which the primordial
interaction volume expands, the energy density falling toward the
critical density of the QCD hadronization point.
I propose that bulk hadron production data
hold the promise to elucidate the conditions prevailing at the
phase transition. In Sect. 4 the implications of strangeness
production data will be discussed. Sect. 5 presents an analysis of
 hadronic yield ratios that are shown to refer directly to
the conditions prevailing at the hadronization point. In Sect. 6 I
turn to hadronic  spectra and Bose-Einstein correlation data
pointing out their potential information content regarding the
overall partonic-hadronic dynamical trajectory of the interaction
volume. Sect. 7 gives a short summary.

\noindent
\section{Transverse Energy Density Estimates}
The prediction of lattice QCD puts the phase transition
between hadrons and partons at about 1.5 $GeV/fm^3$, not
implausible as this is the energy density in the deep interior
sections of hadrons where partons are similarly deconfined albeit
in a small volume. It is the first task of relativistic nuclear
collisions to demonstrate that energy densities upward of 1.5 $GeV/fm^3$
are indeed created in central collisions. To this end one measures
the rapidity distribution of total transverse energy production in
calorimeters, then to relate the rapidity density $dE_T/dy$ to the spatial
density $\epsilon$ in a formalism developed by Bjorken
\cite{2}:
\begin{equation}
\epsilon\:=\:[dE_T/dy]\:[\pi R^2 l]^{-1}.
\end{equation}

\noindent
Employing the Bjorken estimate with a primordial radius
 $R(^{208}Pb)=1.16 \cdot A^{1/3}$, and a formation time $l=1
 fm/c$, the NA49 calorimetric results for the energy density near
 midrapidity were \cite{3}: $\epsilon$=1.3
 (central $S+S$ at $200\:GeV$) and 3.2 (central $Pb+Pb$ at $158\:GeV$)
  (in $GeV/fm^3$),
 i.e.
 \begin{equation}
 \epsilon (Pb+Pb)\approx 2.5\epsilon (S+S).
 \end{equation}

\noindent
In this estimate taking the Woods-Saxon-radius of $Pb$
instead of the $rms$ radius (smaller by about $\sqrt{2}$) leads to
a low value, but taking $l=1fm/c$ (could be up to $l=2\:fm/c$) to a
high value. The combined choices thus appear reasonable.

\noindent
At QM 96 both Kharzeev, and Blaizot and Ollitrault \cite{4} have
pointed out that this is the {\it average} density of the source volume.
The {\it central} energy density is higher by perhaps 1.5
depending on one's picture of the radial density profile. $Pb+Pb$
may thus reach about $4.0 - 4.5\: GeV/fm$ in the extended interior
sections. But the ratio $\epsilon(Pb)/\epsilon (S)\approx 2.5$
independent of these considerations. Thus the "interior" of $S+S$
(if existing) will provide about 1.6 $GeV/fm^3$.
The Bjorken estimate refers to an ultra-relativistic collision
scenario supposing a boost invariant, longitudinally expanding
tube of partonic matter. At the modest SPS $\sqrt{s}\approx20\: GeV$
we are, not quite, at this ideal limit. The domain of approximate
boost invariance shrinks to a relatively narrow interval in
longitudinal phase space, $\mid y_{cm} \mid\: \le \:1$. We focus on
that interval in turning from global, calorimetric $E_T$ data to
energy density estimates based on the momentum space distributions
of identified hadrons.
From NA49 charged hadron production 4 $\pi$  data, G\"unther \cite{5}
gets $\epsilon =2.16 \:GeV/fm^3$, for pions + net baryons
 in central $Pb+Pb$ near mid-rapidity.
Estimate the additional kaon + newly created baryon-antibaryon
fraction to add about 15-20\%:
\begin{equation}
\epsilon(Pb+Pb)_{y\approx y_{mid}}=(2.6 \pm 0.3)\: GeV/fm^3,
\end{equation}
 not so bad an agreement between tracking and calorimetry.
We may obtain a third estimate from the observations that

\noindent
1. The distribution of net baryons is quasi-flat in $ 2<y<4$ and

\noindent
2. The net "proton" m$_T$-spectral slopes are quasi-constant in
$3<y<4$.

\noindent
These NA49 data \cite{5,6} are illustrated in $Fig.~2$.

\begin{figure}
\hspace{0.5cm}
\begin{minipage}[t]{5cm}
\epsfig{file=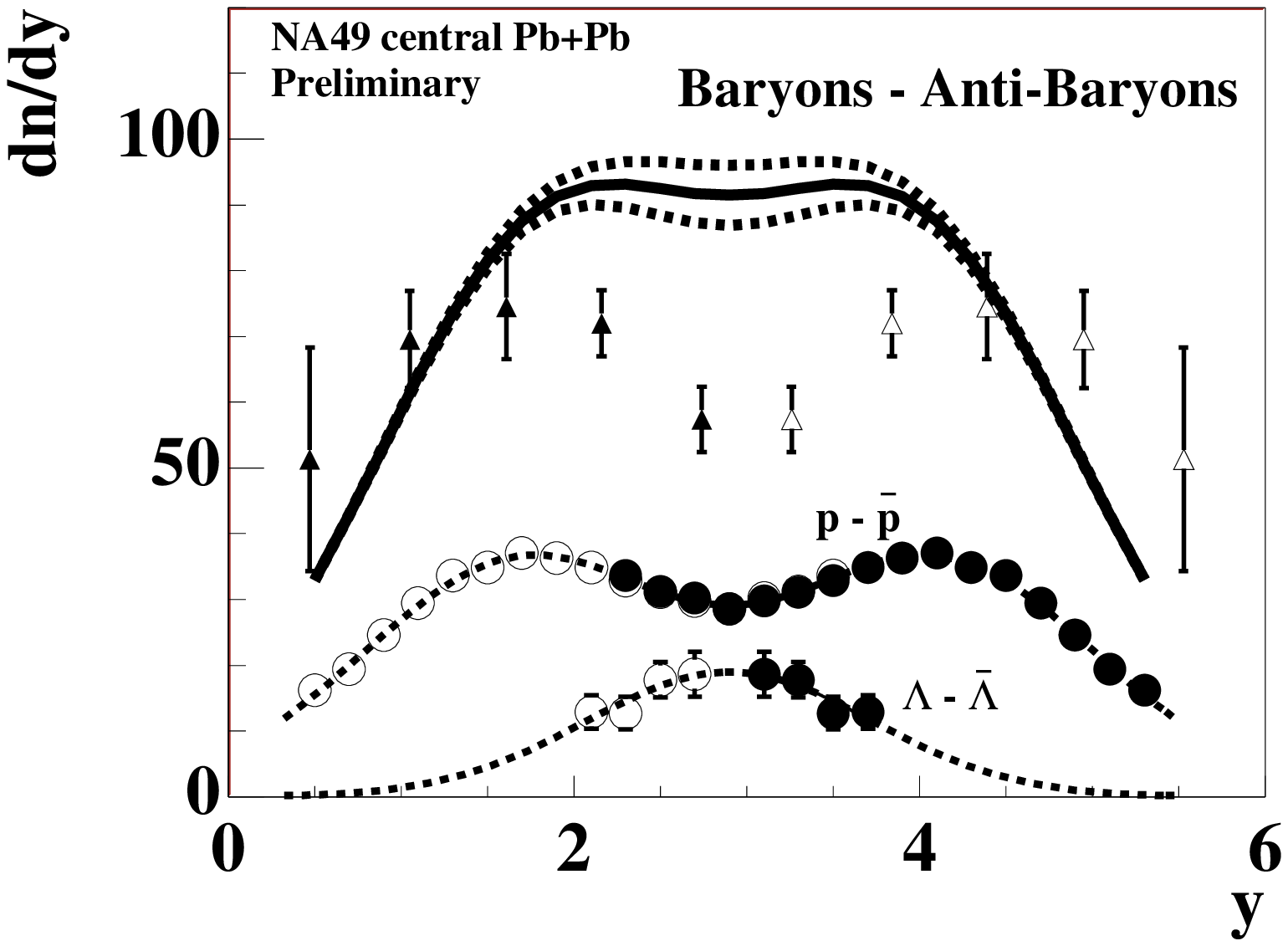,height=6.5cm,width=9cm}
\end{minipage}
\hspace{2.5cm}
\begin{minipage}[t]{5cm}
  \epsfig{file=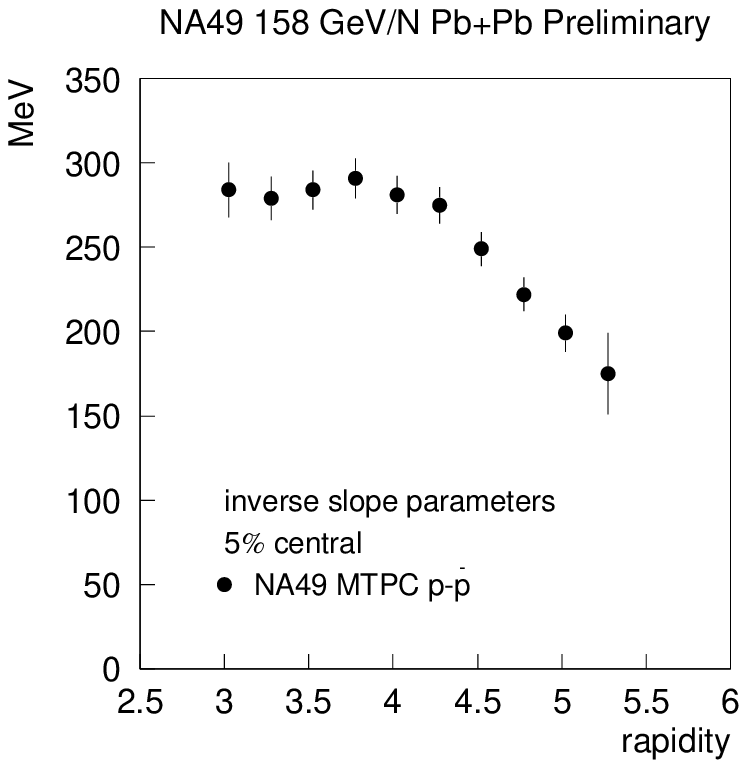,height=7cm,width=8.5cm}
  \end{minipage}
  \vspace{-0.5cm}
  \caption{Rapidity distribution of participant baryons in central
  $S+S$ (triangles scaled up by 7) and $Pb+Pb$
  collisions (left), and inverse slopes of proton transverse mass
  distribution as a function of rapidity (right) \cite{5,6}.}
\end{figure}

\noindent
From this we may infer that the primordial hadron or parton source
is, in rough approximation, a cylinder with $Pb$-radius extending
from $y=2$ to $4$. Assume further that this cylinder "radiates" the
entire transverse energy content of the system, as seen in NA49
calorimeter data \cite{3}:  $E^{tot}_T=(1.0 \pm 0.1)\:TeV$, for central
$Pb+Pb$.

\noindent
Now make a Bjorken-like estimate for the spatial volume of that
cylinder: assume $<R (Pb$ Woods-Saxon)$>=6 fm$ at $<b>=2.0 \:fm$ (the NA49
centrality
trigger \cite{3}); assume that one rapidity unit corresponds to a longitudinal
extension equal to the formation length for which I now take
$l=1.5\:fm$:
\begin{equation}
V_{cylinder}=\pi R^2l \cdot \Delta y = 340\: fm^3
\end{equation}
\begin{equation}
\epsilon_{average} = (2.9 \pm 0.3)\:  GeV/fm^3
\end{equation}
in central $Pb+Pb$. Overall (from the 3 estimates) we get for the
average
\begin{equation}
\epsilon(Pb)=(2.9 \pm 0.4)\:GeV/fm^3.
\end{equation}

\noindent
In summary: Kharzeev, Blaizot and  Ollitrault \cite{4} may be right in
esti\-mating the {\it interior} $Pb+Pb$ energy density to be about $4\:
GeV/fm^3$. It would then be about $1.6\: GeV/fm^3$ in $S+S$. Thus
the latter system may just approach the critical QCD energy
whereas all the above estimates indicate that the fireball created
at mid-rapidity in central $Pb+Pb$ collisions exhibits energy
density beyond the realm of matter consisting of hadrons.

\section{Lattice QCD and Debeye-Screening of $J/\Psi$ and $\Psi'$}

From QCD we need to recall two predictions: First, look at recent
estimates of the critical energy density at which the parton
$\leftrightarrow$ hadron phase change occurs. It has continuously
come down over the last 5 years, to \cite{1}
\begin{equation}
\epsilon^{crit}(QCD \:lattice) = (1.5 \pm 0.5)\: GeV/fm^3.
\end{equation}

\noindent
F. Karsch \cite{7} even gets $1\:GeV/fm^3$.
Second, however, it is important to recall here the earlier
estimates \cite{8} of the QCD "Debeye screening"
length (i.e. the length scale at which QCD acquires an effective
short range interaction form) as a function of energy density.
More than a decade ago Matsui and Satz \cite{8} argued that QCD
bound states (hadrons) should dissolve once their radius exceeds
the screening length. They pointed out the small radius $J/\Psi$
vector meson as a suitable tracer hadron, to monitor QGP plasma
conditions atteined in nuclear collisions.
\begin{figure}
\begin{center}
\begin{minipage}[t]{4cm}
\epsfig{file=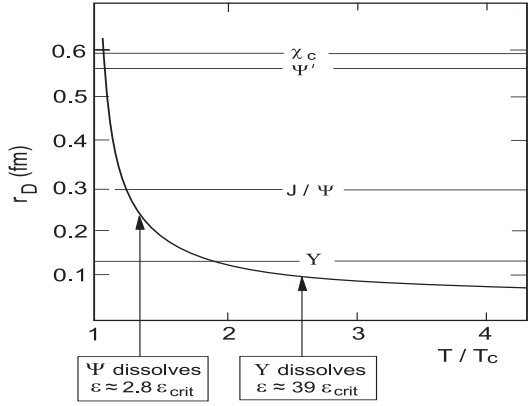,height=6cm,width=6cm}
\end{minipage}
\hspace{5cm}
\begin{minipage}[b]{5cm}
\vspace{-0.5cm}
\caption{Debeye screening radius from QCD \cite{8} versus temperature in
units of $T_c$.}
\end{minipage}
\end{center}
\end{figure}

\noindent
 $Fig.~2$ shows results presented in the book "Quark Matter I" edited
by R. Hwa \cite{9}:
\noindent
The screening length $r_D$ falls steeply with $T/T_c$. For the large
$\Psi'$ we see $r_D\:\approx$ $r(\Psi')$ already at $T=T_c$, but
 $r(J/\Psi) \approx 0.5\:r\:(\Psi')$ and thus screening
 (disruption) of $J/\Psi$ occurs at $T/T_c \approx 1.3$. This
 seems little difference but recall the plasma energy density $\epsilon$
 proportional $T^4$ to first order:

thus $\Psi'$ melts at $T_c$ at which
 $\epsilon=\epsilon^{crit} \approx 1.5\:GeV/fm^3$

but $J/\Psi$ melts at $1.3\:T_c$,
 $\epsilon=2.86\:\epsilon^{crit} \approx 4.3\:GeV/fm^3$.

\noindent
 Our above estimates of energy density thus lead us to expect that
 $\Psi'$ yields are suppressed already in intermediate mass
 collisions, whereas the $J/\Psi$
yield gets critically suppressed in central $Pb+Pb$ collisions
only. This expectation is in fact borne out by
NA38/50 data for $\Psi'$ and $J/\Psi$  suppression \cite{10}.
NA38 reported $\Psi'$ suppression to be "complete" already in
semi-central $S+W$ (roughly comparable to central $S+S$).
NA50 reported $J/\Psi$ suppression to start becoming "complete" in central $Pb+Pb$
only (we ignore here the details \cite{11} of the much-discussed dependence,
differentially, on system size or $E_T$ scales in NA38/50 data).
These data are illustrated in $Fig.~4$.

\noindent
With "complete" I refer to "maximum possible"
suppression: the yield can never go to zero because of the
unavoidable surface regions, at low $\epsilon$,
present in all collision geometries! A provocative conclusion results:
From lattice QCD $\Psi'$ suppression says we are at (or slightly above) $T_c$
and at (or slightly above) $\epsilon^{crit} \approx 1.5 \:
GeV/fm^3$. Indeed NA35/49 estimates \cite{3} for central $S+S$ or semi-peripheral
$S+A=200$ : $\epsilon \approx 1.6\:GeV/fm^3$, and NA38 sees
that $\Psi'$ disappears here. This agreement, by itself, presents
no strong argument as breakup of the large, weakly bound $\Psi'$
could also occur in hadronic matter at this energy density.
However the onset of a disappearing $J/\Psi$ yield in central $Pb+Pb$
($Fig.~4$) signals that we are near $\epsilon = 4\:GeV/fm^3$ here,
in agreement with the above estimates of $\epsilon$. Invoking the
effect of hadronic co-movers is completely implausible here
because the forbiddingly high packing density (several "hadrons"
per $fm^3$) renders hadronic transport or cascade models
meaningless. We conclude that these observations point to the
existence of a non-hadronic phase.

\begin{figure}
\hspace{2cm}
\begin{minipage}[b]{5cm}
\epsfig{file=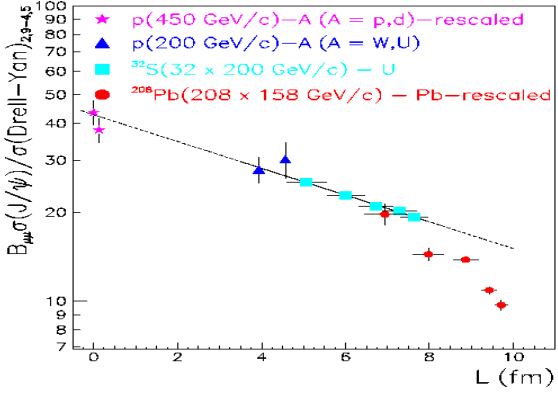,height=4cm,width=5cm}
\end{minipage}
\hspace{1cm}
\begin{minipage}[b]{8.5cm}
 \caption{$J/\Psi$ production relative to the Drell-Yan yield.
  Data for proton, sulphur and lead induced collisions are shown from NA38,50,51.
  They are plotted against the effective thickness of the collision system.
  All data fall onto the $6\:mb$ breakup line \protect\cite{12} except for
  the central $Pb+Pb$ collisions which exhibit additional suppression.}
\end{minipage}
\end{figure}

\section{Total Strangeness Yield and Strange/Entropy Ratio}
From now on I will tentatively take for granted the above
indications
that in central $S+S$ we are at (or slightly above) $T_c$ whereas with
increasing system size we end up at about $1.3\:T_c$ in central
$Pb+Pb$, the interior energy densities ranging up to  about
$4 \:GeV/fm^3$, where a partonic phase is realized. Are other observed signals compatible with this hypothesis?
I turn now to bulk hadron production data to show that this is indeed the
case. Let us first recall the
NA35/49 results concerning strangeness production.

\noindent
The abundance of $s+\overline{s}$ relative to $u+\overline{u}+d+\overline{d}$
at or near hadronization can be estimated by "Wroblewski"
quark-counting \cite{13}; this estimate
starts from the observed strange to nonstrange hadron production ratio in $4\pi$.
It can be approximated by the measured $K/\pi$ and $\Lambda/\pi$
ratios \cite{14}. The  NA35/49 data show that the $K/\pi$ ratio stays
{\it near constant} in central $S+S,\: S+Ag/Au$ and $Pb+Pb$ \cite{6,14,15}.
The yields and yield ratios of strange and nonstrange hadrons in $S+S$
\cite{14} and $Pb+Pb$ \cite{6} exhibit a near perfect hadro-chemical
equipartition in phase space. The analysis by Becattini et al.
\cite{16} shows that all final yields (extrapolating to early times by
inclusion of all resonance decays a la Wroblewski) resemble a
thermal "family" of hadrons at $180<T<190\:MeV$,
for all SPS reactions from $S+S$ to $Pb+Pb$. {\it However this family
can only be consistently described by making the additional
assumption that the strangeness content is
universally underpopulated}, at 60-70\% only of the global
equilibrium abundance in a hadronic "reactor vessel" at these
temperatures. Essentially no change from $S+S$ to $Pb+Pb$, like in
the $K/\pi$ ratio. These results are illustrated in $Fig.~5$.
\begin{figure}
\begin{center}
\epsfig{file=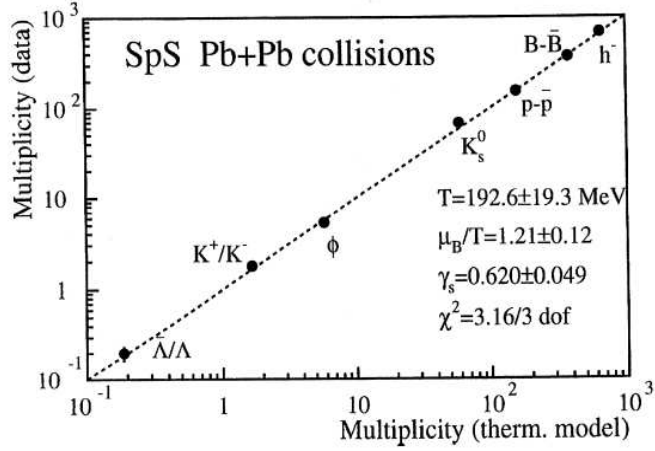,height=9cm,width=9cm}
\vspace{-3cm}
  \caption{Comparison of hadronic yields and yield ratios in central
   $Pb+Pb$ collisions with predictions of the thermal model
\cite{16}. The parameter $\gamma _s$ represents the degree of
strangeness saturation.}
\end{center}
\end{figure}
\noindent
In order to build a conclusion on these observations
 consider the model formulated by Kapusta
and Mekjian \cite{17}.
 They derive estimates for the
dynamical equilibration (relaxation) times of quark flavours in a
model quark gluon gas and in a hadronic reactor, and they derive predictions
for the equilibrium abundance ratios of $K/\pi$ (an
observable that they link to the "strangeness to entropy ratio").
\begin{figure}
\hspace{1cm}
\begin{minipage}[t]{6.5cm}
\epsfig{file=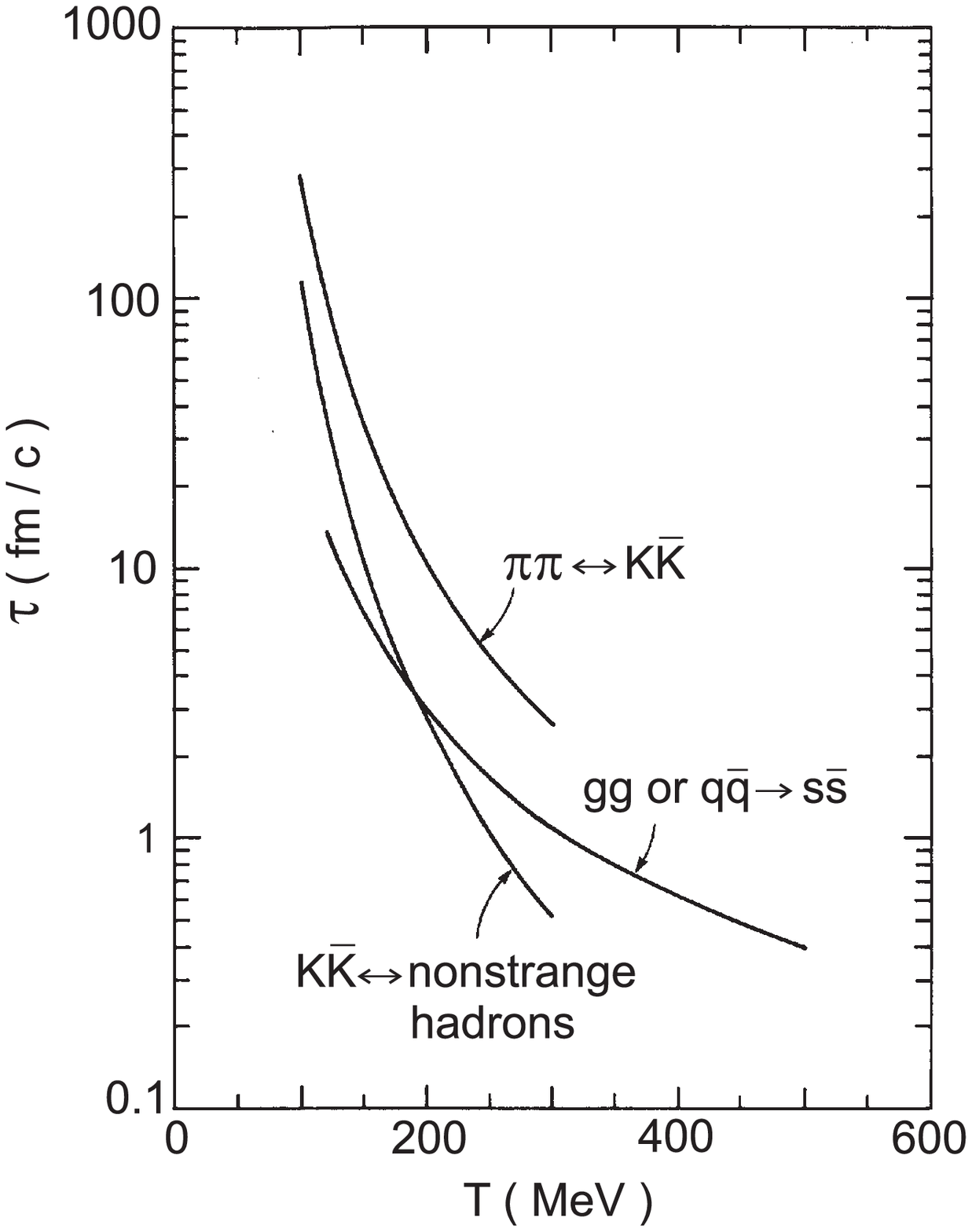,height=6cm,width=6cm}
\caption{Relaxation time constants \cite{16} for hadrons and
partons as a function of temperature}.
\end{minipage}
\hspace{2cm}
\begin{minipage}[t]{7.5cm}
\epsfig{file=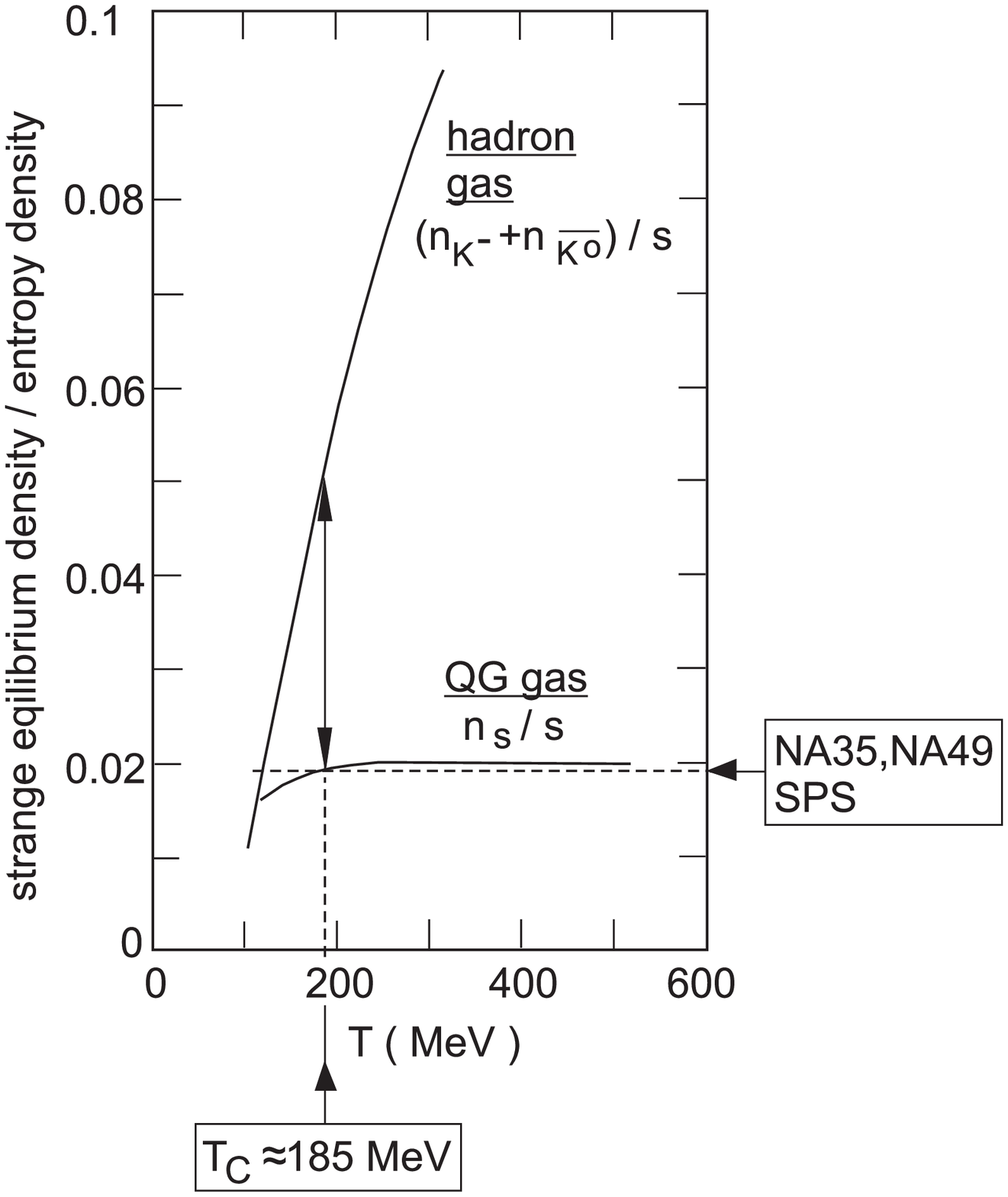,height=6.5cm,width=6.5cm}
\caption{Equilibrium ratio of strangeness to
entropy \cite{16} in a partonic and hadronic scenario}
\end{minipage}
\end{figure}

\noindent
$Fig.~6$ shows their estimates for the equilibration time constant
vs. temperature, in a QGP and in a hadron gas. We assume now  that
the critical temperature is $T_c=185 \pm 15\:MeV$, and
that, in central $Pb+Pb$, we are starting from a primordial state at $T\approx
1.3\:T_c=240\:MeV$. The relaxation time is of order $\tau=1\:fm/c$
in both systems at the latter temperature: comparable to Bjorkens formation
time! The strange to nonstrange content of the reactor vessel must therefore
be near equilibrium at the instant of particle formation under
such circumstances, both for a hadronic or partonic state. $Fig.~7$
shows the ratio of strangeness  to
entropy abundance, essentially translating into the $K/\pi$ ratio.
If the primordial state at $T=240\:MeV$ was in the hadronic phase, the
equilibrium ratio would be about 3.5 times higher than in the
partonic phase, and a huge $s+\overline{s}$ population would
result.Upon expansion we reach $T=T_c\approx 185\:MeV$
where the population would still exceed the QGP population ratio
by a factor of about two. Note that the latter ratio is constant!
No variations of $K/\pi$ are seen in the data, which are near a value
 of 0.15 for all systems and agree with the
"QGP" level \cite{18}.
Furthermore, $Fig.~6$ shows that the relaxation time
$\tau > \: 3 \: fm/c$ below $T=185\:MeV$, rapidly increa\-sing with
temperature falling further: it appears unlikely that the observed
ratio $K/\pi$ is much altered by anything that happens below
$T=185\:MeV$. Indeed the Becattini model ($Fig.~5$) finds
$T=190\:\pm\:20\:MeV$.

\noindent
We are lead to the conclusion that the system was not in a hadronic phase at its
maximum energy density, neither in $S+S$ nor, of course, in
$Pb+Pb$ because it would then have no reason to be strangeness-undersaturated.
If it was in a partonic phase (represented in $Fig.~7$ by the
thermal parton equilibrium state "QG-gas") its $K/\pi$ yield would
be constant throughout, and near the value observed by NA49 as
Sollfrank et al. have shown \cite{18}. As the "QG-gas" hadronizes at $T=185\:MeV$
it does not bring sufficient strangeness into the emerging hadron
phase which, if in global flavour equilibrium by itself would
feature about twice the strangeness content from $Fig.~7$. However
$Fig.~6$ indicates that the hadronic phase would take upward of $3\:fm/c$
at constant $T \approx 185\:MeV$ to equilibrate strangeness. This time
is not available owing to
the rapid expansion prevailing at hadronization time.
The hadronic phase will, thus, evolve essentially preserving its pre-hadronic
strangeness input.
The NA35/49 strangeness data thus appear to agree qualitatively
with expectations from a simple model for a parton to hadron phase transition
occuring at $T\approx 185\:MeV$ from $S+S$ to $Pb+Pb$ \cite{18}.

\noindent
Note that in these thermal models [14,16,17,18] one employs
hadronization temperatures of up to $190\:MeV$
without wondering about the classical Hagedorn limit for the
hadronic temperature, of about $165\:MeV$.
Future models that incorporate the hadronic eigenvolume may thus drastically
change our views \cite{19}. Furthermore, hadrons need not to be in their
vacuum configurations \cite{20} at the high hadronic densities prevailing at
hadronization (contrary to what is assumed in all hadro-chemical models).
These observations keep us, for the present
time, from a firm claim that the strangeness to entropy data imply
discovery of the QCD partonic phase. They are, however, compatible with this
hypothesis.

\noindent
I also wish to note here that the term "thermal model" refers to
statistical phase space descriptions that differ in detail. The
Becattini model \cite{16,18} maintains the strangeness saturation
parameter $\gamma_s$ in keeping with the Wroblewski analysis
\cite{13}
of hadronic collisions where strangeness is manifestly
undersaturated ($\gamma_s = 0.3-0.4)$. For $Pb+Pb$ this model
still suggests a significant deviation from unity ($\gamma_s
= 0.62$): primordial strangeness is enhanced relative to $p+p$, $p+A$
and $e^++e^-$ collisions at similar
energy but still significantly undersaturated in a hypothetical global
equilibrium state at the hadronic side of the phase transition.
$Fig.~8$ illustrates these observations in the framework of the
thermal model \cite{16,18}. They are cast here into the variable
suggested by Wroblewski \cite{13}: strange to non-strange quark
abundance at the state at hadronization.
\begin{figure}
\begin{center}
\epsfig{file=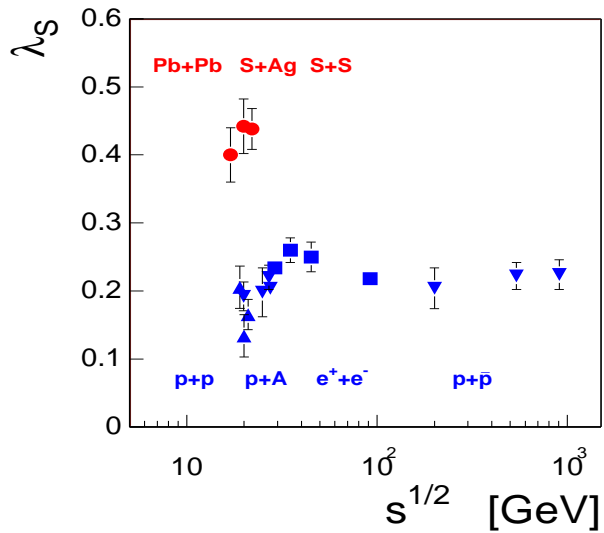,height=6.5cm,width=7.5cm}
\end{center}
\vspace{-1cm}
\caption{The ratio of strange to non-strange quarks,
$\lambda_s=(s+\overline{s})/2(u+\overline{u}+d+\overline{d})$ in
nucleus-nucleus collisions and elementary collisions \cite{16}.}
\end{figure}

\noindent
In the next section we shall suggest that the hadronic population
ratios are the result not of a rescattering equilibrium state created
in the early hadronic phase but of the hadronization mechanism.
Strangeness undersaturation results from the partonic
strangeness levels, in this picture, and central nuclear collisions
exhibit drastic differences from, e.g. $e^+e^-$ regarding their
primordial strangeness population, as seen in $Fig.~8$. However, other thermal
models for SPS collisions dispense with the additional
parameter of strangeness saturation \cite{21} still obtaining reasonable fits to the
hadronic production ratios. The eventual decision concerning strangeness
as a diagnostic of the flavour composition prior to hadronization
will come from new precision data \cite{22} on hyperon production
($\Lambda (1520), \Xi, \Omega$). In fact Bialas \cite{23} has just
shown that the observed hyperon to antihyperon yield ratios can be
well understood in a flavour coalescence model (see next section)
based on the partonic strangeness concentration.

\section{Hadronization and apparent Hadrochemical Equilibrium}
\subsection{Origins of Equilibrium}
The observation of an apparent thermal equilibrium among hadronic species
has baffled particle physicists since Hagedorn's times. As I have
made ample reference to such models above I insert a section
to attempt a qualitative explanation, in a scetchy manner.
Let me note, first, that chemical equilibrium among a mixed phase
of inelastically interacting hadronic species represents, not at
all a difficult situation but the maximum entropy state (minimum
information): the state of {\it highest} statistical probability.
In a multiparticle inelastic collision far above thresholds the key
question must be (opposite to the traditional approach) what could
keep the system from realizing that state. We will show in sec. 6
that the very fast, "explosive" expansion mode observed at SPS energy
indeed prohibits hadro-chemical equilibration by  rescattering.
Nevertheless the concept of chemical equilibration as a limit of multiple
inelastic rescattering cascades is the presently most persued
point of view. The dominance of this view is due to theoretical
implementation in microscopic cascade models which have shown that, at
modestly relativistic energy, the lowest modes of hadronic matter
can indeed acquire equilibrium due to
rescattering in central heavy nucleus collisions \cite{24}.

\noindent
However this "rescattering paradigma" must lead to deep skepticism
concerning equilibrium concepts once the collision energy
increases (dominance of longitudinal motion), from $\sqrt{s} \approx 2.5\:GeV$
at Bevalac/SIS to $\sqrt{s} = 20\:GeV$ at the SPS. Furthermore a hadronic
rescattering mechanism is obviously inapplicable once we turn to
individual hadron or lepton collisions. The apparent success of
Hagedorn analysis at $\sqrt{s} \ge 8\:GeV$ would, thus remain a mystery.
However this analysis turns out to be even more satisfactory once
the energy increases, and $e^+e^-$ LEP collider data for $Z^0$ decay
to hadrons at $92\:GeV$ exhibit perfect equilibrium populations of about 20
hadronic species \cite{25}. The temperature from the fit is $190\:MeV$,
the same as in $Pb+Pb$ ($Fig.~5$)! But the strangeness
undersaturation factor is more prominent, $\gamma_s=0.4$.
Obviously these observations are incompatible with the hadronic
rescattering paradigma.

\noindent
There must thus be another way to create a hadrochemical
equilibrium state at $T = 185 \:MeV$. The answer has been
indicated by Geiger and Ellis
\cite {26}: the {\it hadronization process} enforces phase space
dominance due to its combined non-perturbative mechanisms.
This view has first been presented by Knoll et al. \cite{31}
Geiger and Ellis studied similar LEP data as Becattini \cite{16},
$W^{+-}$ to hadrons, in a partonic
transport model which ends in hadronization. The latter is treated
as a multiple "coalescence" in which the right combinations of
partonic spin, flavour and colour are combining to form colour
neutral
pre-hadrons (heavy resonances that decay instantaneously). The
observed hadron production yields are well accounted for,
and the authors note a remarkable insensitivity regarding the
detailed assumptions made for the hadronization mechanisms. The
final multihadronic state is thus {\it born into equilibrium} (i.e.
at maximal entropy), out of the partonic phase.

\noindent
A tantalizing conclusion results. If SPS $Pb+Pb$ central
collisions create a partonic initial phase we should observe
similar hadronic population ratios as in the LEP data, as a consequence
of the system evolving through a parton to hadron phase
transition. This is indeed the case ($Fig.~5$ for $Pb+Pb$)! In this
view, the increase in $\gamma _s$ from 0.4 to 0.62 and the upward
jump $\lambda_s$ ($Fig.~8)$ {\it results
from differences in the partonic phase}, between the $e^+e^-$
single initial "string" and the large transverse dimension in the
nuclear collision. No new strangeness is created in the
Geiger-Ellis model, in the process of hadronization.
\begin{figure}
\begin{center}
\epsfig{file=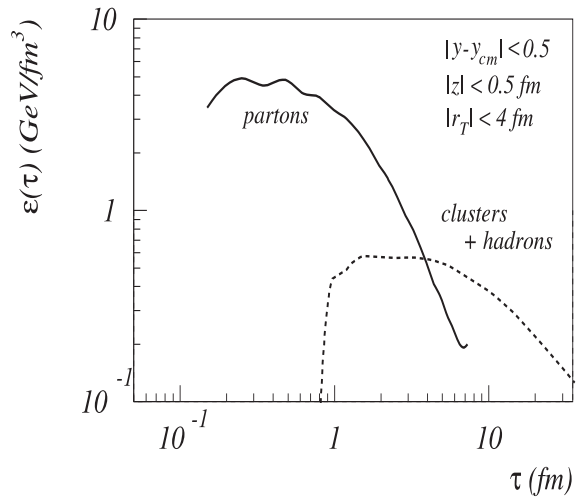,height=6.5cm,width=7.5cm}
\end{center}
\vspace{-1cm}
\caption{Energy density for central $Pb+Pb$ at SPS from the parton
  transport and hadronization model of Geiger and Srivastava \cite{27}.}
\end{figure}

\noindent
Geiger and Srivastava \cite{27} have recently been daring enough to apply
this model to SPS central $Pb+Pb$ collisions. They conclude
that there are remaining remnants of
projectile-target structure functions but that the hadron yield
near midrapidity stems mostly from a parton cascade to
hadronization process. In fact they are able to reproduce the proton
to negative hadron to kaon ratios reported by NA49 \cite{6}. Of particular
interest in view of our discussion in sections 2 and 3 is their
result for the overall time dependence of the energy density in an
interior subvolume of $4\:fm$ transverse extension near midrapidity,
as reproduced in $Fig.~9$. The partonic phase exhibits $\epsilon > 2\:GeV/fm^3$
until about $\tau=2\:fm/c$; hadronization begins after a formation
time of about one $fm/c$ and the hadronic phase ends
at $\tau \approx 20\:fm/c$ where $\epsilon<0.2\:GeV/fm^3$. The
initial energy density amounts to $4\:GeV/fm^3$. All this agrees
with our above guesses. In particular there are no hadronic
co-movers to a $c \overline{c}$ pair of any significant density,
as to break up the emerging $J/\Psi$ signal.
The hadronic fraction never exceeds an energy density of $0.5\:GeV/fm^3.$
Note that this model does thus not predict an unreasonably high
hadronic density during the hadronization phase, apparently due to
the long duration ($4-5\:fm/c$) of the hadronization transition:
hadrons can escape in transverse direction, creating a radial
expansion velocity pattern (see next section).

\noindent
We conclude that hadronic equilibrium populations at temperatures
as high as $185\:MeV$ can be understood as a fingerprint of
QCD hadronization. In this view the multihadronic final state is
not the result of cascade inelasticity. The rigid fixation
of $T \approx 185\:MeV$ conditions throughout the reaction
volume (which must have significant primordial variation of energy density
due to impact geometry) stems from the universal avenue
through hadronization which occurs at a rigidly fixed energy density.

\subsection{Analysis of Single Events}
If the above line of argument is correct we have thus located the
QCD phase boundary at $T\:\approx\:185\:MeV$ for a baryochemical
potential of $\mu_B\:\approx\:0.25\:GeV$, specific for the
conditions reached in central $S$ and $Pb$ induced collisions at
top SPS energy. From similar analysis of multi-hadronic final
states created in LEP $Z^0$ decays we infer that the transition
"temperature" is about  the same at $\mu_B=0$. This indicates
a {\it universal} influence of the non perturbative QCD hadronization
mechanism at sufficiently high $\sqrt{s}$, and at low values of
$\mu_B$. What, then, is specific to central nuclear collisions?
Let me proceed in two steps.

\noindent
First, there can be no doubt that an $e^+e^-$ generated $Z^0$
decay "string" of $92\:GeV$ constitutes an ideal QCD excitation
object free of net quantum number constraints and structure
function remnants, clearly of strictly partonic composition, which
must hadronize and, thus, reveal QCD hadronization features. The
emphasis in the line of argument in sections 4 and 5 is to
demonstrate that $Pb+Pb$ at $\sqrt{s}=17\:GeV$ shows similar
hadronization features and might thus also result from a
partonic phase. This is a totally non-trivial proposition as the
elementary baryon collisions at this $\sqrt{s}$ should have an
average parton-parton $\sqrt{s}$ of about $3\:GeV$ only, and the
dynamics is beset by quantum number conservation constraints.  It
is thus very hard to conceive that all substructures of the
initial baryons should be wiped out and melted into a primitive
uniform partonic phase. However, whereas the primary $Z^0$ decay
"string" is of dramatic longitudinal extension (of about $90\:fm$)
but of small transverse size directly decaying into the vacuum,
the short "strings" of primordial baryonic collisions (of about $3\:fm$
length) remain trapped in a large radius cylindrical collision
volume of modest aspect ratio (as shown in section 2). The vacuum
might be expelled to an outer surface far remote from each primordial
string. At the prevailing energy density of several $GeV/fm^3$ the
"string" substructure (precarious anyhow at
$\sqrt{s}\approx3\:GeV$) must melt away similarly to the initial
baryon structure as the duration of the high density phase (see
$Fig.~9$) far exceeds the mean decay time of a free QCD string.
The specific significance of the SPS $Pb+Pb$ data concerning
hadronic yield ratios thus goes beyond reflecting the fingerprint
of the QCD hadronization mechanism (remarkable enough at
$\sqrt{s}\approx17\:GeV)$: it seems to indicate that bulk partonic
volumes hadronize similarly to isolated longitudinal "strings".

\noindent
Secondly, however, we observe differences in detail that point to
the specific features of a large coherent partonic fireball, in
comparison to a thin string. We have seen specific suppressions
$(J/\Psi$)
and enhancements (strangeness to entropy ratio) in sections 3 and
4. Also, recall $Fig.~8$. An important final hint, supporting
the above picture of a bulk
parton to hadron phase transition, may be derived from recent  NA49
data concerning the {\it event by event} fluctuation of the $K/\pi$
ratio in central $Pb+Pb$ collisions \cite{6}. $Fig.~10$ shows that
this quantity (which indicates both the strangeness to entropy
ratio and the overall hadro-chemical makeup of the final state)
exhibits no fluctuations other than inflicted
by counting statistics as the histogram of single event ratios is
nearly identical to the signal derived from artificially Monte Carlo
generated mixed events.

\noindent
Translating this result we may conclude that all central $Pb+Pb$
collision events are identical as to their thermodynamical
properties, the observed histogram resulting from sampling
statistics only. This observation is made possible by the large
acceptance of NA49. It helps to reject, first of all, the
critical argument raised oftentimes against thermodynamical
analysis, namely that taking ensemble averages fakes thermal
patterns which are not a property of individual events. To the
contrary these data show that each event permits canonical
analysis like a small but macroscopic thermodynamical system, in
line with our intention to study partonic or hadronic {\it bulk
matter} properties in such collision events. More specifically (in
view of our above discussion) these data support the hypothesis
that the apparent hadro-chemical equilibrium state is not caused by
hadronic rescattering cascades which would probably lead to a
larger dispersion (from the dense interior to the dilute surface
regions) as far as the $K/\pi$ ratio is concerned - and thus to a
broad distribution toward  lower ratios. On the other
hand, if the system reaches above the critical energy density
of $\epsilon \approx 1.5\:GeV/fm^3$ in most of its volume, the
dynamical prehistory gets wiped out because both the much denser
central sections and the still dense-enough outer sections of the
fireball uniformly encounter QCD hadronization at a fixed energy
density (corresponding to $T\approx 185\:MeV$) albeit at different
hadronization times. The primordial spread in geometry and
dynamics thus reflects in a hadronization {\it time} spread
(c.f.$Fig.~8$), accessible to Bose-Einstein correlation
study (see next section). However the outcome of hadronization
seems to be thus common over the entire fireball volume, in each
event: this could explain the data in $Fig.~10$.

\begin{figure}
\begin{minipage}[t]{6cm}
\hspace{4cm}
 \epsfig{file=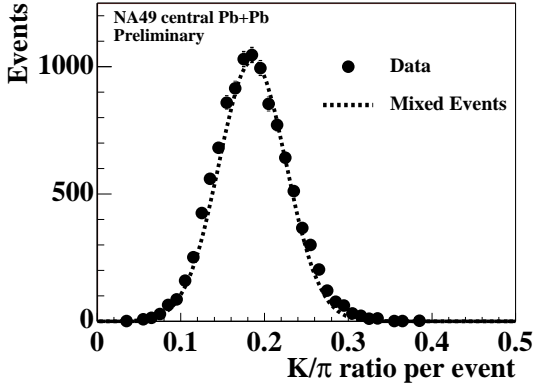, width=6cm}
 \end{minipage}
\hspace{4.5cm}
\begin{minipage}[t]{6cm}
\caption{Event by event fluctuation of the $K/\pi$ ratio in central
$Pb+Pb$ collisions in the domain $3.5<y<5$ as reported by NA49 \cite{6}.
This ratio is obtained from summing negative and positive kaons and
pions. The absolute scale is arbitrary (no acceptance corrections
applied). The dashed line refers to artificial mixed events.}
\end{minipage}
\end{figure}

\noindent
\section{\Large Hadronic Expansion Dynamics}
Now to the final point: the dynamics of the hadronic expansion. From a
combined study of $\pi \pi$ Bose-Einstein (HBT) correlation and final
hadronic $m_T$ spectra we learn, first of all (NA49 ref. [28], NA44
ref.[29]),that
hadronic expansion is different in $S+S$ and $Pb+Pb$. The
transverse energy increases drastically \cite{6,29}, the geometrical HBT
parameters increase drastically \cite{28}. Two different classes
of hadronic observables are of relevance here. First one observes
a breaking of the $m_T$ scaling behaviour predicted in
a simple fireball model for the inverse slope "temperatures" (all
hadronic species should exhibit similar transverse mass spectral
slopes).
\begin{figure}
\begin{minipage}[t]{6cm}
\hspace{4cm}
 \epsfig{file=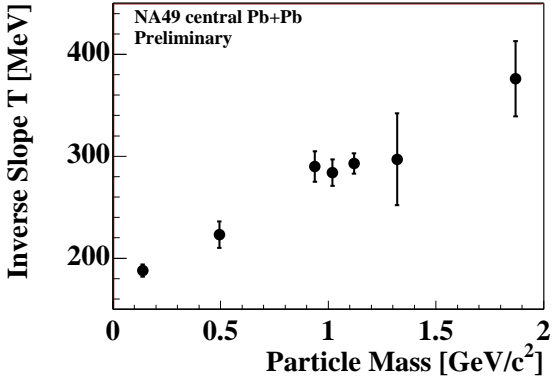,width=6cm}
 \end{minipage}
\hspace{4.5cm}
\begin{minipage}[t]{6cm}
\caption{Increase of the inverse slope parameter $T$ of transverse
mass spectra with the hadronic mass, from $\pi$ to deuterons, in central
$Pb+Pb$ collisions \cite{6}.}
\end{minipage}
\end{figure}
$Fig.~11$ shows NA49 results \cite{6} for $m_T$ spectral
slopes in $Pb+Pb$ collisions, for hadrons from $\pi$
to deuterium: the spectral slope parameter increases from 180 to
380 $MeV$. Similar data have been reported by NA44 \cite{29}.
This behaviour has been linked to a collective
transverse velocity field \cite{29}, prevailing in the
expansion. This field "blue-shifts" transverse mass spectra in
order of hadronic mass. The final hadronic spectra thus result
from superposition of a thermal velocity spectrum, corresponding
to the true temperature of hadrons decoupling from strong
interaction (freeze-out), and from a radial velocity field
blue-shifting that temperature. The origin of the collective
velocity field must reside in the overall dynamics of expansion,
prior to freeze-out. The spectral data, alone by themselves, do
not provide for a clear-cut separation of the two superimposed
effects. However, two pion Bose-Einstein correlation (HBT) data
allow for an independent analysis of the two combined effects.
Combining these two sources of information \cite{28} leads to determination of
the thermal, and collective velocity field ingredients of the
freeze-out stage. $Fig.~12$ shows the result of this analysis (for
detail the reader is referred to \cite {28} and references
therein). The "true" temperature at the end of all strong
interaction is about $T \approx 120\:MeV$, and the system has
developed a collective, radially symmetric transverse velocity
profile with $\beta_{\bot} \approx 0.6$ at the freezeout-hypersurface.

\begin{figure}
\hspace{1cm}
\begin{minipage}[b]{5cm}
\epsfig{file=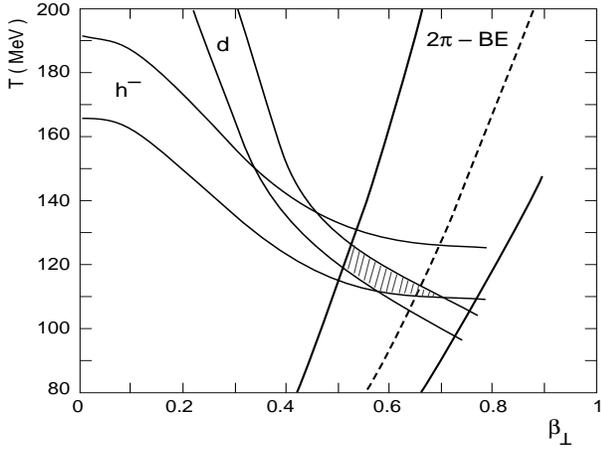,height=6cm,width=8cm}
\end{minipage}
\hspace{4cm}
\begin{minipage}[b]{7cm}
  \caption{Allowed regions in the plane of freeze-out temperature
  and transverse collective velocity derived \cite{28} from negative
  hadron and deuteron $m_T$ spectra, as well as from Bose-Einstein
  two pion correlation in central $Pb+Pb$ collisions.}
\end{minipage}
\end{figure}

\noindent
From the space-time HBT parameters, the $m_T$ spectra and from our
previous arguments about the hadronization temperature, equal to
$T_c$, we may give the following picture of a central $Pb+Pb$
collision:
\begin{enumerate}
\item The system originates at about $T=240\:MeV$ and $\epsilon=4\:GeV/fm^3$
in the interior
and triples its volume until it arrives at $T_c \approx
180-190\:MeV$, $E_c \approx 1.5\:GeV/fm^3$. In a spherical
approximation this would take about 4-5 $fm/c$
if the universal expansion velocity was $\beta \approx 0.5$. In
reality the faster longitudinal expansion will shorten this time
to perhaps $2\:fm/c$ \cite {27} (c.f. $Fig.~9$).
\item It hadronizes at $T$ ("Becattini") $\approx 185\:MeV$at
various baryochemical potentials depending on the reaction system,
$S+S$ to $Pb+Pb$, but near $\mu_B=0.25 \:GeV$. Of course hadronization
does not occur instantaneously throughout  the
volume. It may take 4 $fm/c$ \cite {27}.
\item During expansion from primordial conditions the pion pair
emission strength is
represented \cite{28}
by a Gaussian of mean (overall life-time of the source) 8$fm/c$ and
sigma  (duration of emission time) $4\:fm/c$: this is shown in $Fig.~13$. Pion pair
emission starts right after the formation time of about $1 \:fm/c$
from the surface of the system. I.e. the overall "life-time"
starts at this time. The luminosity peaks at about $8\:fm/c$.
These features are in agreement with the
Geiger and Srivastava predictions ($Fig.~9)$.

\item During this interval, it grows in transverse and
longitudinal directions. The transverse $rms$ radius \cite{28} increases by a
factor of about 2.5. The transverse density thus falls by a factor
of 6.25 in a time interval of about $8\:fm/c$: we observe an
"explosive" expansion pattern.

\item It freezes out from strong interaction with  $T \approx 120
\:MeV$, the freeze-out phase ending  at $\tau \approx 15\:fm/c$.
Collective transverse
and longitudinal velocity fields are observed with Gaussian mean
velocities at the $rms$
points of the density profiles $\beta_{\bot} \approx 0.55$ and
$\eta_{\parallel} \approx 0.9$ \cite{28}.
\end{enumerate}

\noindent
It is our expectation that this set of data will so severely
constrain dynamic expansion models that the conditions
at hadronization will be pinned down independent of all other
 information.
\begin{figure}
\vspace{1cm}
\begin{center}
  \epsfig{file=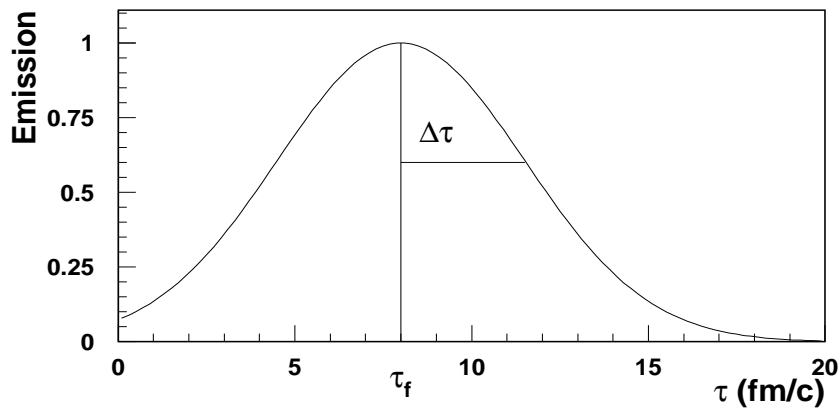,height=5cm,width=5cm}
   \caption{The time profile of pion pair emission from the interaction
  volume: overall Gaussian life-time $\tau_f$ and duration of
   emission parameter $\Delta\tau$ for central $Pb+Pb$ collisions
   from Bose-Einstein correlation analysis \cite{28}.}
  \end{center}
\end{figure}

\section{Discussion}

\noindent
The essential, new point of view in the line of argument persued here
stems
from revision of our understanding of the observed hadronic
production ratios that suggest $T \approx 185\:MeV$ in central $Pb+Pb$
collisions
similar to LEP $Z^0$ and $W$ decays to hadrons. We propose this apparent
hadronic  equilibrium state to
be a fingerprint of the non-perturbative hadronization process
following Ellis and Geiger \cite{26} .
Nuclear collisions at SPS energy reflect the same hadronization
properties, which appear not to arise from  chemical equilibrium
attainment by inelastic rescattering cascades.
The change of mechanisms is highlighted by  the fact that hadronic
rescattering can at SPS energy not even alter the $T \approx 185\:MeV$
abundance pattern throughout hadronic expansion (due
to very fast, "explosive" expansion). Its observation in central
nuclear collisions thus lends support to the existence of a transient
partonic phase that enters non-perturbative hadronization, after some initial
expansion. The implied quantum number coalescence mechanism can be
directly tested in hyperon to antihyperon production
ratios \cite{23,30}. Very recent such data \cite{22} by WA97 and NA49
 appear to support the evidence for a parton to hadron
transition by statistical flavour coalescence.

\noindent
In summary, I have tried to demonstrate that the majority of CERN SPS data can
be coherently understood by assuming that the reaction dynamics of
central collisions
reaches beyond the hadronic phase throughout the reactions studied
yet, i. e. $S+(S,\:Ag,\:Au,\:Pb,\:W,\:U)$ and $Pb+Pb$. The
primordial energy density in sulphur beam reactions may be just at
or above the critical energy density $\epsilon_c \approx
1.5\:GeV/fm^3$ whereas central $Pb+Pb$ collisions should promote the
primordial energy density to far above $\epsilon_c$; we
estimate from $J/\Psi$ suppression data and calorimetry that the
density in the extended interior sections of the reaction
volume  reaches ca. $4\:GeV/fm^3$, clearly beyond the realm of
hadronic matter. For clarity of argument I note that such a
primordial passage into a partonic scenario occurs in a
non-equilibrium process. It remains to be further investigated
whether expansion time scales, partonic relaxation times etc.
conspire favourably for the system to approach the equilibrium QCD state
"quark-gluon-plasma" - the object of desire - before expansion
brings it back to hadronization. However, bulk hadron production
data appear to fix the latter to occur in the vicinity of about
$180\:MeV$: the phase boundary has thus been tentatively located.

\vspace{1cm}
\noindent
$^*$  This article is devoted to the memory of Klaus Geiger.
Presented at the Erice School of Nuclear Physics, Sept. 1998. To
be published in {\it Progress in Particle and  Nuclear Physics}.

\end{document}